\documentclass[a4paper,fleqn,usenatbib]{mnras}

\usepackage[T1]{fontenc}
\usepackage{ae,aecompl}

\usepackage{graphicx}	
\usepackage{amsmath}	
\usepackage{amssymb}	
\hypersetup{draft}

\title[The formation of \text{HAT-P-26b}]{Possible formation pathways for the low density Neptune-mass planet \text{HAT-P-26b}}

\author[M. Ali-Dib \& G. Lakhlani]{
Mohamad Ali-Dib$^{1}$$^{,}$$^{2}$\thanks{E-mail: m.alidib@utoronto.ca}
and Gunjan Lakhlani$^{2}$$^{,}$$^{3}$
\\
$^{1}$Centre for Planetary Sciences, Department of Physical \& Environmental Sciences, University of Toronto at Scarborough,\\
Toronto, ON M1C 1A4, Canada\\
$^{2}$Canadian Institute for Theoretical Astrophysics, 60 St. George St, University of Toronto, Toronto, ON M5S 3H8, Canada\\
$^{3}$Department of Physics, University of Toronto, Toronto, ON M5S 1A7, Canada
}

\date{Accepted XXX. Received YYY; in original form ZZZ}

\pubyear{2016}

\begin{document}
\label{firstpage}
\pagerange{\pageref{firstpage}--\pageref{lastpage}}
\maketitle

\begin{abstract}
We investigate possible pathways for the formation of the low density Neptune-mass planet HAT-P-26b. We use two formation different models based on pebbles and planetesimals accretion, and includes gas accretion, disk migration and simple photoevaporation. The models tracks{ the atmospheric oxygen abundance, in addition to the orbital period, and mass} of the forming planets, that we compare to HAT-P-26b. We find that pebbles accretion can explain this planet more naturally than planetesimals accretion that fails completely unless we artificially enhance the disk metallicity significantly. Pebble accretion models can reproduce HAT-P-26b with either a high initial core mass and low amount of {envelope enrichment through core erosion or pebbles dissolution}, or the opposite, with both scenarios being possible. {Assuming a low {envelope enrichment factor} as expected from convection theory and comparable to the values we can infer from the D/H measurements in Uranus and Neptune, our most probable formation pathway for HAT-P-26b is through pebble accretion starting around 10 AU early in the disk's lifetime.}
\end{abstract}

\begin{keywords}
planets and satellites: formation -- planets and satellites: gaseous planets -- planets and satellites: composition 
\end{keywords}


\section{Introduction}
Neptune-mass planets are among the most abundant in the galaxy \citep{winn, mullaly}. These are thought to be an intermediate step in giant planets formation, being planets massive enough to accrete significant amounts of gas, but not able to undergo runaway gas accretion before the dissipation of the disk \citep{pollack1996, benz}. They hence offer a direct glimpse into the formation processes of both low and high mass planets. The benchmarks for this planetary category are Uranus and Neptune, ice giants with metallicities (measured through methane abundance) $\sim$ 100 $\times$ solar \citep{baines,karko, cavalie}. Recently, a Neptune mass planet (HAT-P-26b) was observed with a relatively low bulk density, and a metallicity (measured through water abundance) upper limit of 30 $\times$ solar \citep{hatp, hartman}. This is in contrast with HAT-P-11b, another Neptune-mass planet where the metallicity was consistent within the large error bars with Uranus and Neptune \citep{hat11}. This poses the challenge of explaining the density diversity among Neptune-mass planets. While multiple works focused on the gas giants \citep{mousis} and Uranus and Neptune \citep{robinson,ali-diba, helled}, this paper aims to understand the origin of the low metallicity of HAT-P-26b. The question this work tries to answer is hence:\\
\textit{Is it possible to form a low metallicity Neptune-mass planet on short orbit using current planets formation paradigms ?}\\
Using {end-to-end} planets formation models based on pebbles and planetesimals accretion, we offer different formation scenarios for HAT-P-26b, with the goal of constraining its possible formation location and time, {and its connection to Uranus and Neptune}. Moreover, we also want to understand if low density Neptune-mass planets accreted a small {amount of heavy elements} to start with, or a significant core that remained decoupled from the envelope due to weak core erosion.

\section{Model}
We use in this work the model from \cite{ali-dibc, ali-dibd}, based on \cite{bitsch1,lamb1}. {It includes fits to a 2D disk model as the backbone, on top of which we add parametrized pebbles accretion for both the Bondi (where planetesimals accretion is initially important) and Hill accretion regimes. Pebble accretion in the Hill regime can proceed in either ``2D'' (when the planet's Hill radius is larger than the scale height of the pebbles disk) or a less efficient ``3D'' manner in the other case around. {The switch from the Bondi to the Hill regime happens when the approach speed of the pebble becomes dominated by the Hill speed instead of the sub-Keplerian speed.}}

{We hence define the pebbles accretion rate in the Hill regime as:
\begin{equation}
\label{p2d}
\dot{M}_{c,peb,2D} = 2 \bigg(\frac{\tau_f}{0.1}   \bigg)^{2/3} r_H v_H \Sigma_{peb}
\end{equation}
and:
\begin{equation}
\label{p3d}
\dot{M}_{c,peb,3D} \propto \dot{M}_{c,peb,2D} \times \frac{r_H}{H_{peb}}
\end{equation}
{where we notice that optimally coupled pebbles (with a Stokes number $\tau_f \sim 0.1$) are accreted efficiently from all of the planet's Hill radii, thus making this channel much faster than planetesimals accretion.} We refer the reader to \cite{ali-dibd} for a discussion on the less efficient Bondi accretion regime for smaller protoplanets. The pebbles scale height $H_{peb}$ is defined as: }

\begin{equation}
H_{peb} = H_g \sqrt{\alpha / \tau_f} 
\end{equation}

{where $\alpha$ is the disk's turbulence parameter fixed to $\sim 5 \times 10^{-3}$. }

{In classical planetesimals accretion models, core growth proceeds till reaching the critical mass (around 10 M$_\oplus$) where hydrodynamic gas accretion (collapse) starts. In pebble accretion models however, the much higher solids accretion rates lead to significantly higher envelope luminosity, and thus critical masses in the order of 100 M$_\oplus$ \citep{lamb2}, which is incompatible with observations. An important concept in pebble accretion models is hence the pebble isolation mass, defined as the mass over which the core will change the disk's gas pressure profile around it, thus deviating pebbles instead of accreting them. This allows for a natural cut-off for the core accretion phase, decreasing the luminosity and allowing for gas accretion to proceed. 
The model therefore includes pebbles isolation mass as:
\begin{equation}
M_{iso} \sim 20 \bigg( \frac{H/r}{0.05} \bigg)^{3} M_\oplus
\end{equation}

in addition to the slow (hydrostatic growth) and fast (hydrodynamic collapse) phases gas accretion (for respectively M$_{core} >$ M$_{envelope}$ and the opposite case. } It also includes type I and II migration through torques formula evaluation (including the corotation torque and its saturation), and finally simplified disk photoevaporation limiting the lifetime of the disk and affecting its metallicity.
Moreover, the model follows the oxygen abundance of the forming planets, allowing us to evaluate their metallicity as tracked by this observable. We follow \cite{ali-dibc} in dividing the disk abundance of oxygen into refractories (in solid phase at all temperatures) and the volatile H$_2$O (in solid phase only outside the snowline). This ratio is chosen to be consistent with cometary values. The planet's core mass and total mass of heavy elements are also readily obtained from the model.

{Both of Uranus and Neptune have high envelope metallicities measured through methane. This enrichment can have multiple origins:
\begin{itemize}
\item Core {erosion} into the envelope after the formation of the planet. It is not clear however if actual {core erosion} can occur in Neptune-mass planets. This would necessitate either the core being gradually mixed into a metallic-hydrogen layer, or a vigorous convection at the core-envelope boundary \citep{guillot2004, wilsonmilitzer,vazan}. Both planets are not likely to have metallic-Hydrogen layers, and while Neptune's interior in thought to be convective, it might not be the case for Uranus \citep{guillot1994, podolak1995}. Even if convection is present in Neptune-mass planets, it is probably not as strong as in Jupiter.
\item The accretion of metals-heavy gas due to the photoevaporation of the disk (\citep{guillot2006} and as included in our model).
\item The dissolution of the accreted solids in the envelope during the formation of the planet (before the accretion of significant amounts of gas, \citep{podolak})
\item The planet's envelope pollution with external planetesimals after it finished accreting, but we do not take this effect into account as it was demonstrated to be very ineffective \citep{gladman}.
\end{itemize}
Now we argue that for the purpose of our paper, 1 and 3 lead (to first order) {to similar outcomes and can be modelled simultaneously using a simple fraction parameter ranging from 0 $\%$ (no {erosion or dissolution}) to 100 $\%$ (full {erosion or dissolution}).} We hence do not distinguish between metals who were part of the core then got eroded into the envelope, and those who were accreted during the first phase of core accretion but fragmented or sublimated before reaching the core. This is because whatever the solids are dissolved \textit{during} the accretion, or eroded from the core \textit{after} the accretion, in both cases we would end up with a residual core (less massive that the ideal case core with no {erosion} or dissolution) and a metals rich envelope. In these two cases the ``{envelope enrichment factor}'' would have different interpretations, where it would imply the fraction of solids that were dissolved into the envelope of the planet before reaching the core, or the fraction that was eroded later for the other case around.  Whichever of these two mechanisms is the main source of the metals enrichment in the atmospheres of Uranus and Neptune is debatable. Both cases however (to first order) will lead to the same rough isolation mass value, and the same fractional distribution of metals between the envelope and the solid core, leading thus to identical observables (metallicity through water abundance). In both cases however we need to assume the envelope to be fully convective, for the observed upper atmosphere water abundance to be representative of the bulk value. The validity of these results (and most current planet formation models), depend on this assumption.}

For a more detailed description of the model we refer the reader to \cite{ali-dibc} and the references therein. The planetesimals accretion model is described in section \ref{plansection}.

{Simulations were run on a discrete grid with 0.01 AU resolution. We inject only one planet per disk, and stop the simulations when a its reaches the inner edge of the grid, or the disk fully photoevaporate. We then only keep the planets with mass between 15 and 21 M$_\oplus$ on orbits shorter than 10 days {(that did not reach the inner edge)}, with an oxygen abundance between 1 and 30 $\times$ the solar value to be consistent with HAT-P-26b.

{All of the used parameters and their values are shown in tables \ref{t1}. We explore the free parameters space through population synthesis, with $\sim 2.25\times 10^5$ simulations in total. {These parameters are: T$_{ini}$, R$_0$ (the seed's injection time and location), {E$_f$ (the envelope's {enrichment factor})}, and $\dot{M}_{FUV}$ (the disk's photoevaporation rate). We fix the value of the following parameters in all simulations: $f$ (a fudge factor that reconciles our simplified slow phase gas accretion rate parametric fit with more detailed hydrodynamic simulations), $\rho_c$ (the core's density, not a free parameter), ``metal'' (the disk's metallicity in small coupled dust grains), and Z$_0$ (the disk's metallicity in large decoupled pebbles) with the disk's total metallicity (metal + Z$_0$) fixed to the solar value as measured in star HAT-P-26. We also fix the fraction of oxygen in refractory (non volatile rocks) to the value measured in comet 1P/Halley (cf. \cite{ali-dibexo} for a discussion of this parameter).}
We are left with $\kappa_{env}$ (the envelope opacity), and M$_0$ (the seed's initial mass), which are tunable parameters. While we fix their values in the nominal simulations, we explore their effects in section \ref{caveat}.}

\section{Results \& interpretation}
\subsection{Pebbles accretion}
The main knobs controlling the water's abundance in the envelope of a planet in our model are its initial core mass and the amount of {envelope enrichment}. The initial core mass of a planet is defined through the pebbles isolation mass, beyond which the core stops growing and start accreting gas. The isolation mass scales to the cube of the disk's scale height ($\propto (H/r)^3$, \cite{bitsch1}), and thus increases further out in the disk. For a fixed location, it is higher in young disks and decreases with time. {Therefore, in our model, HAT-P-26b can either form late in the inner parts of the disk, acquire a small core, then erode significant amounts of it into the envelope (or dissolve the same amount of solids during accretion), or form early in the outer disk then erode small part of its massive core. We conduct simulations as outlined above to quantify this analysis. }

Results of our simulations are shown in Figure \ref{fig:mcore} showing the oxygen abundance enrichment as a function of the core's isolation mass and {the enrichment factor}. From the initial $2.25 \times 10^5$ simulations, only a few fit all of the constraints we have on HAT-P-26b. The trends seen confirm our analytical discussion where planets with higher isolation mass can reproduce the water's metallicity upper limit observed in HAT-P-26b's envelope only for small enough {envelope enrichment factor} ($\leq 20\%$). Planets with lower isolation mass on the other hand are allowed to have a higher {enrichment factor}s (all the way up to $80\%$). {However we note that, if we assume a total mass of heavy elements in Uranus \& Neptune in the order of 12-16 $M_\oplus$ of mostly ices \citep{helledneptune}, only low {enrichment factor}s less than around 20\% can fit their D/H ratios \citep{ali-dibc}, assuming cometary D/H values for the ices. This is compatible with the discussion above on the low efficiency of convection in Neptune-mass planets.} This implies that if we assume this {enrichment factor} value to be an upper limit on the efficiency of convection (or any other {erosion} mechanism) in this planetary type, and thus this should be similar in HAT-P-26b, then this would favor the cases with high isolation masses and low {enrichment factors}. 

{Additional constraints on HAT-P-26b can be indirectly inferred via interior structure modeling. \cite{hartman} for example used the models of \cite{fortney} to conclude that HAT-P-26b's mass is equally distributed between its core and Hydrogen-Helium envelope . This result is inherently model dependent. We however test now the effects of this constraint on the formation of the planet. Results for this case are shown in Fig. \ref{fig:mcore2}, that's identical to Fig. \ref{fig:mcore} except for this additional constraint. We notice that while a smaller number of simulations now fit this constraint, the distribution of these planets is very similar to the nominal case. This is except the disappearance of the cluster for isolation mass between 8 and 10 M$_\oplus$ seen in the E$_f$ = 0.2 case. This is simply because for these planet the core will dominate the total mass.  }

{In Figure \ref{fig:track} we plot the formation track of a typical HAT-P-26b like planet. The planet will start forming at 10 AU, and accrete pebbles while migrating inward. When reaching the isolation mass, it will stop accreting solids and start contracting gas via the Kelvin–Helmholtz mechanism. The envelope will however never reach the mass of the core before the disk dissipate, and hence the planet will never grow into a gas giant.}

{In Fig. \ref{fig:frequ} we plot the frequency of HAT-P-26b like planets satisfying all of the constraints as a function of where in the disk there seed was initially injected, and thus started forming. This distribution clearly peaks around 10 AU, making the 8-11 AU the optimal region for HAT-P-26b's formation. This is simply due to this region of the disk having aspect ratios values compatible with HAT-P-26b's metallicity, while being at the right distance for migration to bring the planet in to where it is observed today. }

Moreover, in Fig. \ref{fig:mcoreall} we plot the isolation mass of all Neptune-mass planets found by our simulations (with no additional constraints) as a function of their seed's injection time and location in the disk. We notice that, as expected, the planets isolation masses decrease in the inner disks and with time. Hence planets forming early in the outer disk will have a more massive core than those forming late in the inner disk. There is an obvious degeneracy between the 2 parameters controlling the isolation mass.  {If we however assume a 20\% {envelope enrichment factor}, and thus a 10 $M_\oplus$ initial core mass, then HAT-P-26b should have formed early in the disk's lifetime. If we do not assume a preferred {envelope enrichment} value on the other hand, then we cannot remove the degeneracy between the formation location and time of HAT-P-26b.}

We conclude from these simulations that both scenarios are possible for HAT-P-26b. {To be consistent with Uranus and Neptune however within the assumptions made by \cite{ali-dibc}, a relatively massive core with weak {envelope enrichment} is favored. The difference between HAT-P-26b and our own ice giants in this case would be Uranus and Neptune's formation significantly further out in the disk, leading to higher isolation masses for these planets and thus higher metallicities for the same {enrichment factor}s.}

\subsection{Planetesimals accretion}
\label{plansection}
{We now ran simulations where we replace pebbles accretion with the less efficient classical planetesimals accretion.
The basic planetesimals accretion rate following \cite{pollack1996}, for gravitational focusing factor of unity, is:
\begin{equation}
\dot{M}_{c,plan} = \Sigma_{plan} R_c^2 \Omega  
\end{equation}
with the effective capture radius:
\begin{equation}
R_c = r_H \sqrt{\frac{r_c}{r_H}}
\end{equation}
we then get the same equation as \cite{lamb2}:
\begin{equation}
\dot{M}_{c,plan} \simeq \frac{1}{\sqrt{\Phi}} \times r_c \Sigma_{plan} v_H
\end{equation}
where $r_c$ is the core radius, $v_H$ is the Hill speed, and $\Sigma_{plan} = Z \times \Sigma_{\text{gas}} $ is the surface density in planetesimals. We try 2 values for the metallicity $Z$ of 1$\%$ (MMSN) and 8$\%$. We have hence implicitly assumed that the planetesimals dispersion velocity is the Hill velocity. Additionally:
\begin{equation}
\Phi = \frac{r_c}{r_H} \simeq 3\times 10^{-4} \bigg(\frac{r_p}{10 \ \text{AU}}   \bigg)^{-1}
\end{equation}
is the accretion efficiency, showing that planetesimals accretion is up to 4 orders of magnitude less efficient than pebble accretion as discussed below. We multiplied the planetesimals accretion rate by $\frac{1}{\sqrt{\Phi}}$ to account for planetesimals fragmentation into sizes between 0.1 and 1 km \citep{lamb2}, improving their accretion efficiency. This model is hence up to three orders of magnitude more efficient at forming cores than simple cases assuming intact bigger planetesimals, {but still one to two orders slower than pebble accretion (equations \ref{p2d} and \ref{p3d} above, from \cite{lamb2012}).}

The core will continue growing till it reaches the planetesimals accretion isolation mass \citep{bitsch1} :
\begin{multline}
M_{iso,plan} = 0.16 \bigg( \frac{b}{10 R_H}  \bigg )^{3/2}  \bigg( \frac{\Sigma_{plan}}{10}  \bigg )^{3/2}  \bigg( \frac{r}{1 AU}  \bigg )^{1.5(2-s_{pla})} \\
\times \bigg( \frac{M_\star}{M_\odot}  \bigg )^{-0.5} M_E
\end{multline}
where b is the orbital separation of the growing embryos, which we follow \citep{bitsch1} in setting to 10$R_H$, $s_{pla}$ is the negative gradient of the surface density in planetesimals which we set to 0.5, and the stellar mass $M_\star$ which we set to solar value. We start the simulations with a core initial mass of $10^{-4}$ M$_\oplus$, an order of magnitude higher than the value we use for pebbles accretion.\\
We first run simulations with $Z$ = 1$\%$ (MMSN value). {In this case however, this formation channel is too slow, and no planetary cores will grow beyond around 1 M$_\oplus$ (as also noted by \cite{bitsch1}).}
The results remain unchanged even if we change the initial core mass to $10^{-3}$ M$_\oplus$. \\
We now run simulations with $Z$ = 8$\%$ (8 times higher than MMSN value), with an initial core mass of $10^{-5}$ M$_\oplus$. In this case, we do find planets that fit all of the constraints we have on HAT-P-26b. These are shown in Figure \ref{fig:plan2}. Most of these planets cluster around an isolation mass of 7-10 M$_\oplus$, and they are uniformly distributed in the R$_0$ space. They seem to have mostly started forming relatively early (no later than 1 Myr), what allows enough time for the cores to grow through the slow planetesimals accretion. \\
These simulations show that planetesimals accretion can form an HAT-P-26b like planet, but only if we significantly increase the planetesimal metallicity in the disk.

\subsection{Discussions}
\label{caveat}

\subsubsection{Effects of free parameters}
{Here we explore the effects of M$_0$ (the seed initial mass) and $\kappa_{env}$ (the envelope's metallicity) on results. In the pebble accretion model we started the simulations with M$_0$ = $10^{-5} M_{\oplus}$, where the seed will grow initially through the slow planetesimals accretion before reaching the ``pebble transition mass'' around $10^{-3} M_{\oplus}$ and start accreting pebbles \citep{bitsch1}. If we however inject seeds directly with $M_0 = 10^{-3} M_{\oplus}$, those will grow much faster. We would hence expect this case to be atleast as efficient in created HAT-P-26b like planets. This is confirmed by simulations as shown in Fig. \ref{fig:mcore}. For the planetesimals accretion model however, as discussed in section \ref{plansection}, starting with M$_0$ = $10^{-5} M_{\oplus}$ will almost never lead to HAT-P-26b like planets, and starting with more massive seeds is necessary. This is because planetesimals accretion is slower throughout the entire core formation process. \\
The effect of $\kappa_{env}$ on the other hand is to control the gas accretion speed in the initial slow phase, with lower $\kappa_{env}$ accelerating this phase of the formation process. We run simulation with $\kappa_{env} = 0.01$ cm$^2$g$^{-1}$, significantly lower than the nominal $\kappa_{env} = 0.05$ cm$^2$g$^{-1}$ we used. As seen in Fig. \ref{fig:mcore}, lower $\kappa_{env}$ leads to almost no HAT-P-26b like planets. This is because lower envelope opacity will accelerate gas accretion, increasing the efficiency of forming Jupiter-mass planets, at the expense of Neptune-mass planets since those are now much more likely to have enough time to undergo the runaway gas accretion and transition into gas giants. {We note that both opacities we tried are relatively low. We used 0.05 and 0.01 cm$^2$/g for respectively the high and low opacities cases. In comparison, the ISM opacity is around 1 cm$^2$/g. Our opacities are in the same order as the values found by \cite{mov} and \cite{ormel2014}. They are however an order of magnitude higher than the best fit values found by \cite{morda2014}} }

\subsubsection{caveats}
{A major aspect of planet formation our model does not take into account is the feedback effects of the envelope structure evolution during the initial accretion stages on the formation of the planet. \cite{vent} for example included self-consistently the effects of envelope enrichment through sublimation of icy planetesimals with the envelope structure equations and accretion, and concluded that this change in opacity and the mean molecular weight can accelerate the formation of gas giants significantly. 

{On the other hand, \cite{alibert} included pebbles thermodynamics along with the envelope structure equations and the disk advective wind \citep{ormel}. He concluded that pebbles will get dissolved long before hitting the core, even for a 1 M$_\oplus$ planet. Moreover, gas recycling might keep the envelope's metallicity at stellar value, except late in the disk when the planetary Hill radius becomes comparable to the disk scale height.} {Further work however is needed to confirm \cite{alibert} conclusions.}

These works highlight the importance of including such effects into formation models.  However this would necessitate coupling our model to a complete envelope structure model, which is outside of the scope of this work (and would increase the numerical complexity significantly). The absence of a simplified image for the main effects at play in these models also hinders their implementation in population synthesis models.

\section{Summary \& Conclusions}
In this work we investigated the formation of HAT-P-26b like Neptune-mass planets with metallicity below 30 $\times$ solar. We used a planets formation models that includes pebbles or planetesimals accretion, in addition to gas accretion, type I and II migration, and disk photoevaporation. We moreover split the disk's oxygen abundance into refractories and water, and we track its abundance as an indicator of metallicity. Pebble accretion models were efficient enough to form planets that fit all of the constraints, and those were found to have isolation masses between 2 and 10 M$_\oplus$. Lower {enrichment factor}s are needed with high isolation masses to remain below the metallicity upper limit. Both scenarios with high and low initial core mass are thus possible. Planetesimals accretion on the other hand with MMSN-like metallicity was too inefficient to allow the formation of HAT-P-26b, that had to form in the outermost parts of the disk to have an isolation mass high enough, but these could not migrate to inside 10 days before the dissipation of the disk. HAT-P-26b could have formed through planetesimals accretion though if the disk was very rich in metals. Our main conclusion is that the current paradigms of planets formation can account naturally for low density Neptune-mass planets. {The most probable formation pathway for HAT-P-26b from our model would be a planetary embryo forming around 10 AU, early in the disk, to acquire the appropriate core isolation mass of around 10 M$_\oplus$, with at most 20\% in mass of this core ending up in the atmosphere through {erosion} or envelope enrichment during the early formation. This {enrichment factor} is compatible with the values expected in Uranus and Neptune. More observables, like the metallicity through carbon abundance as measured in Neptune, are necessary to make more meaningful comparisons of the formation pathway of HAT-P-26b and our own ice giants.}

\renewcommand\arraystretch{1.2}
\begin{table}
\begin{center}
\caption{Disks and planets parameters space.}
\footnotesize
{\begin{tabular}{lcccc}
\hline
\noalign{\smallskip}

Model parameter			& Range			& Step	\\
\hline
T$_{ini}$			& 10$^5$ - 5$\times$10$^6$		& 10$^5$	 yr	 \\
R$_0$	& 0.5 - 15 AU		& 0.25 AU				 \\		
E$_f$				& 0 - 100\%  	& 20\%	\\		
\hline
$\dot{M}_{FUV}$ (M$_{\odot}$/yr) &  $\mu= 2 \times 10^{-9}$ &	 $\sigma= 2 \times 10^{-9}$ \\
\hline
M$_0$				& 1 $\times$ 10$^{-5}$ M$_\oplus$		& -		\\	
metal & 0.5 \% & - \\
Z$_0$ & 1 \% & -  \\
$f$  & 0.2 & - \\
$\kappa_{env}$ & 0.05 cm$^2$ g$^{-1}$ & - \\
$\rho_c$ & 5.5 g cm$^{-3}$& - \\
Water iceline & 150 K & - \\
Refractory oxygen & 43 \% & - \\
\hline	
& HAT-P-26's parameters &  \\
\hline	
Mass &   0.86 M$_{sun}$ \\
Radius & 0.78 R$_{sun}$ \\
metallicity & 0.01 $\pm 0.04$ dex \\

\hline	
& HAT-P-26b's parameters &  \\
\hline	
Mass &   18.6 $\pm$ 2.2 M$_{Earth}$ \\
Radius & 6.3$^{+0.8}_{-0.4}$ R$_{Earth}$ \\
Orbital period &  4.2 days  \\
eccentricity & 0.12 \\
Water abundance &  $4.8^{+21.5}_{-4.0}$ \ $\times$ solar  \\
\hline	
\end{tabular}}\\

\label{t1}
\end{center}
\end{table}

\begin{figure*}
\begin{centering}

	\includegraphics[scale=0.30]{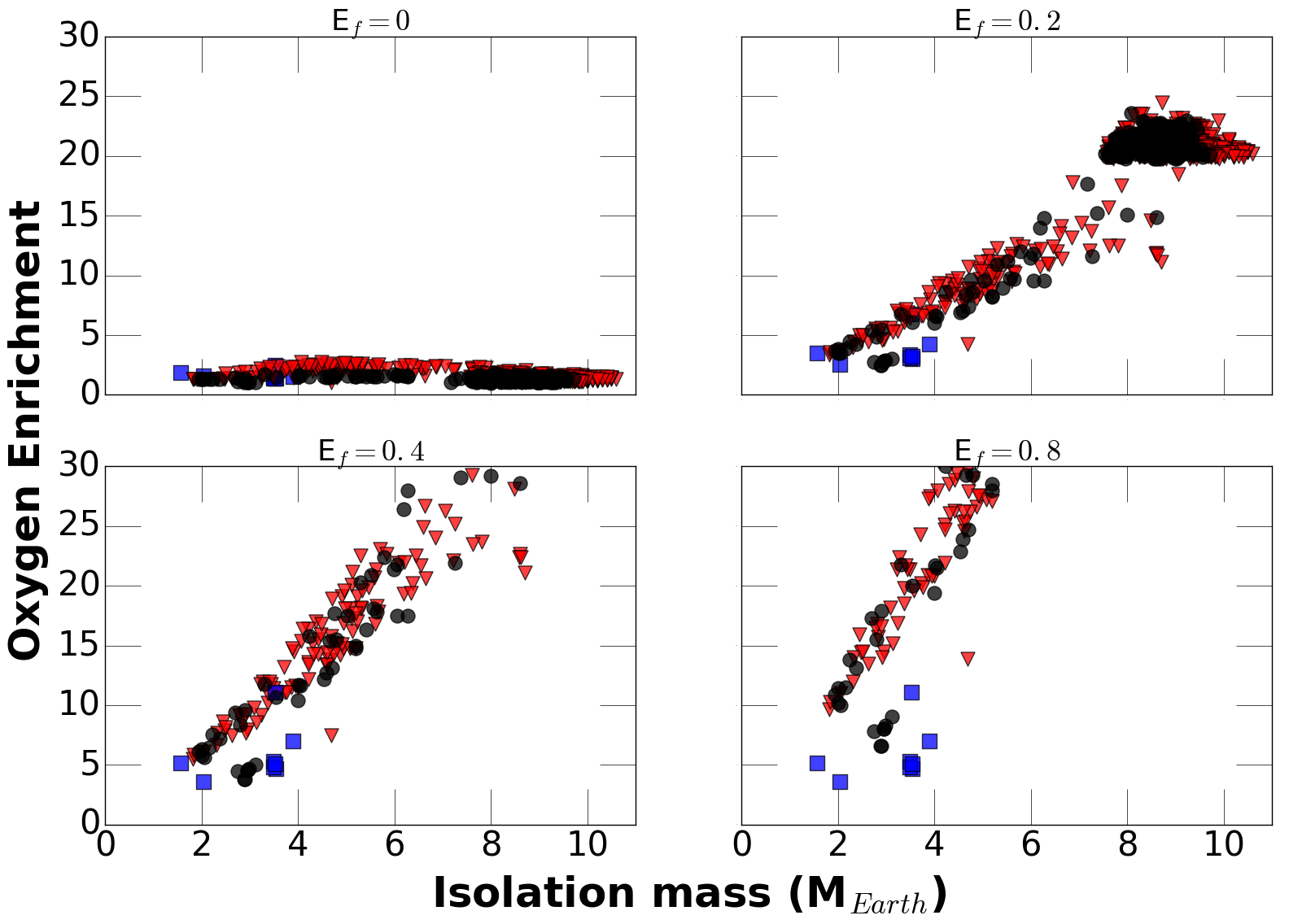}
   \caption{The oxygen enrichment with respect to the solar value as a function of the core's isolation mass, for all of the planets compatible with the constraints of HAT-P-26b. These are hence exclusively Neptune-mass planets that ended up inside 10 days with an oxygen enrichment less than or equal to 30. The 4 subplots correspond to 4 different {envelope enrichment factor}s ($E_f$). {Circles represent our nominal model with $\kappa_{env} = 0.05$ cm$^2$g$^{-1}$ and $M_0 = 10^{-5}$, while squares represent the case with $\kappa_{env} = 0.01$ cm$^2$g$^{-1}$ and triangle the case with $M_0 = 10^{-3}$. We notice two populations fitting HAT-P-26b: planets with low initial core mass and high {enrichment factor}, and planets with high initial core mass and low {enrichment factor}. }}
    \label{fig:mcore}
    \end{centering}
\end{figure*}

\begin{figure*}
\begin{centering}

	\includegraphics[scale=0.30]{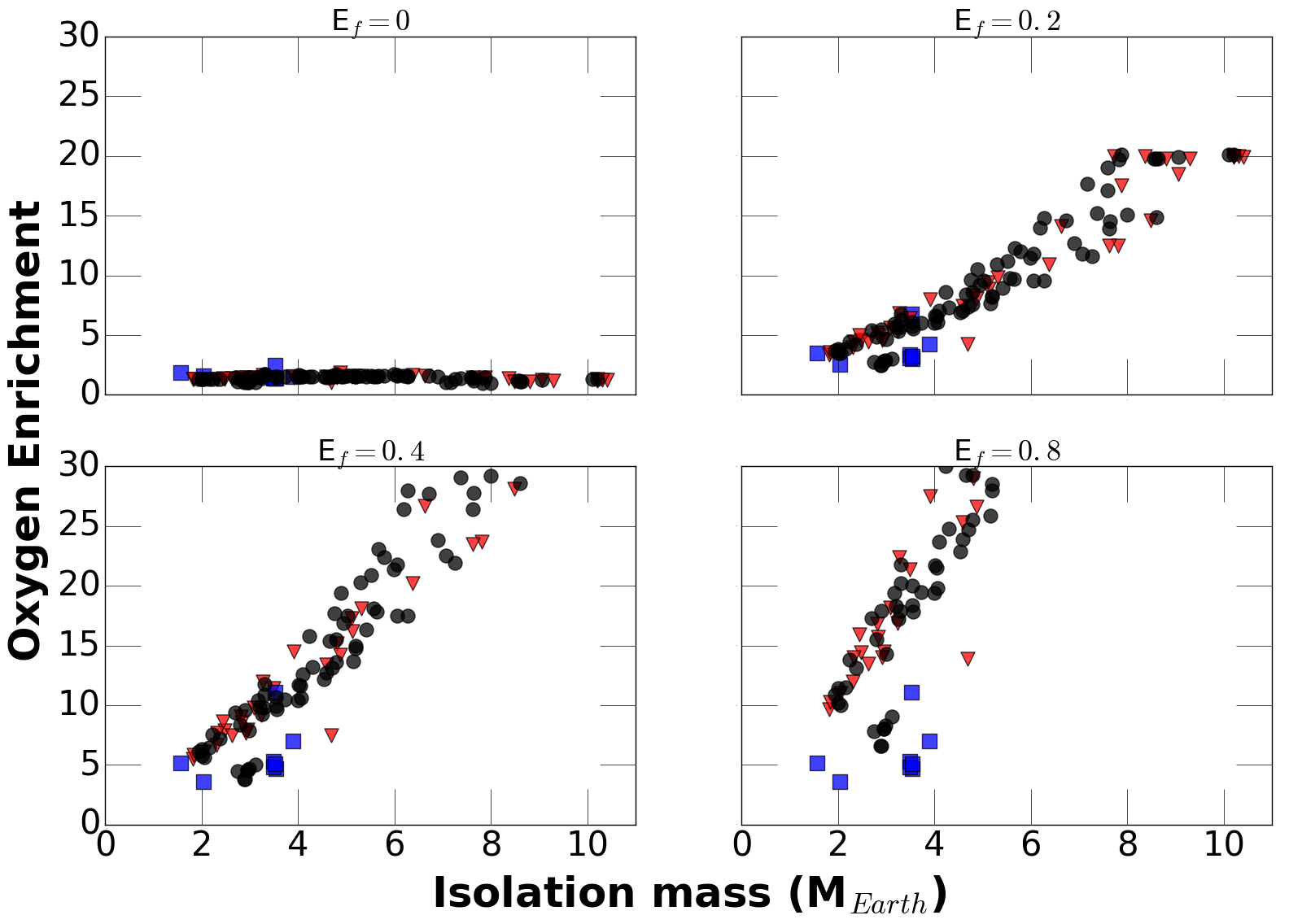}
   \caption{{Same as Fig. \ref{fig:mcore}, but with the additional model dependent constraint that the hydrogen-helium envelope is $\geq$
50 \% of the total mass.}}
    \label{fig:mcore2}
    \end{centering}
\end{figure*}

\begin{figure*}
\begin{centering}
	\includegraphics[scale=0.40]{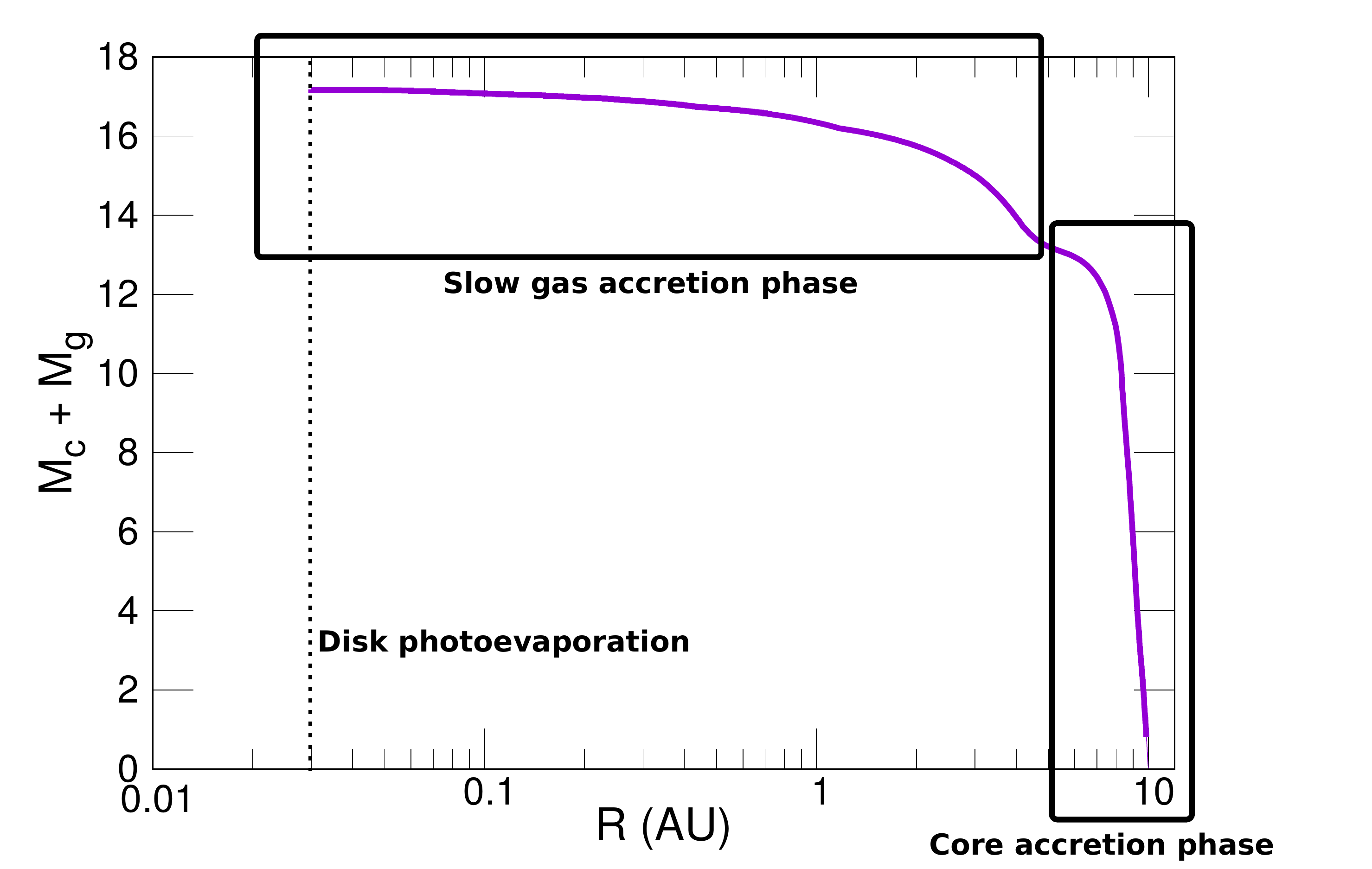}
   \caption{{A formation track for a typical HAT-P-26b like planet in our simulations with pebble accretion, showing the evolution of the total (core and gas) mass as a function of the planet's position. Initially, the seed is injected at 10 AU. It starts accreting solids and move inward through type I migration till reaching the isolation mass. The planet will then start slowly contracting gas and continue migrating inward till the full photoevaporation of the disk. The planets formation is hence stopped during the slow gas accretion phase, before reaching the fast hydrodynamic collapse.}}
    \label{fig:track}
    \end{centering}
\end{figure*}

\begin{figure*}
\begin{centering}
	\includegraphics[scale=0.23]{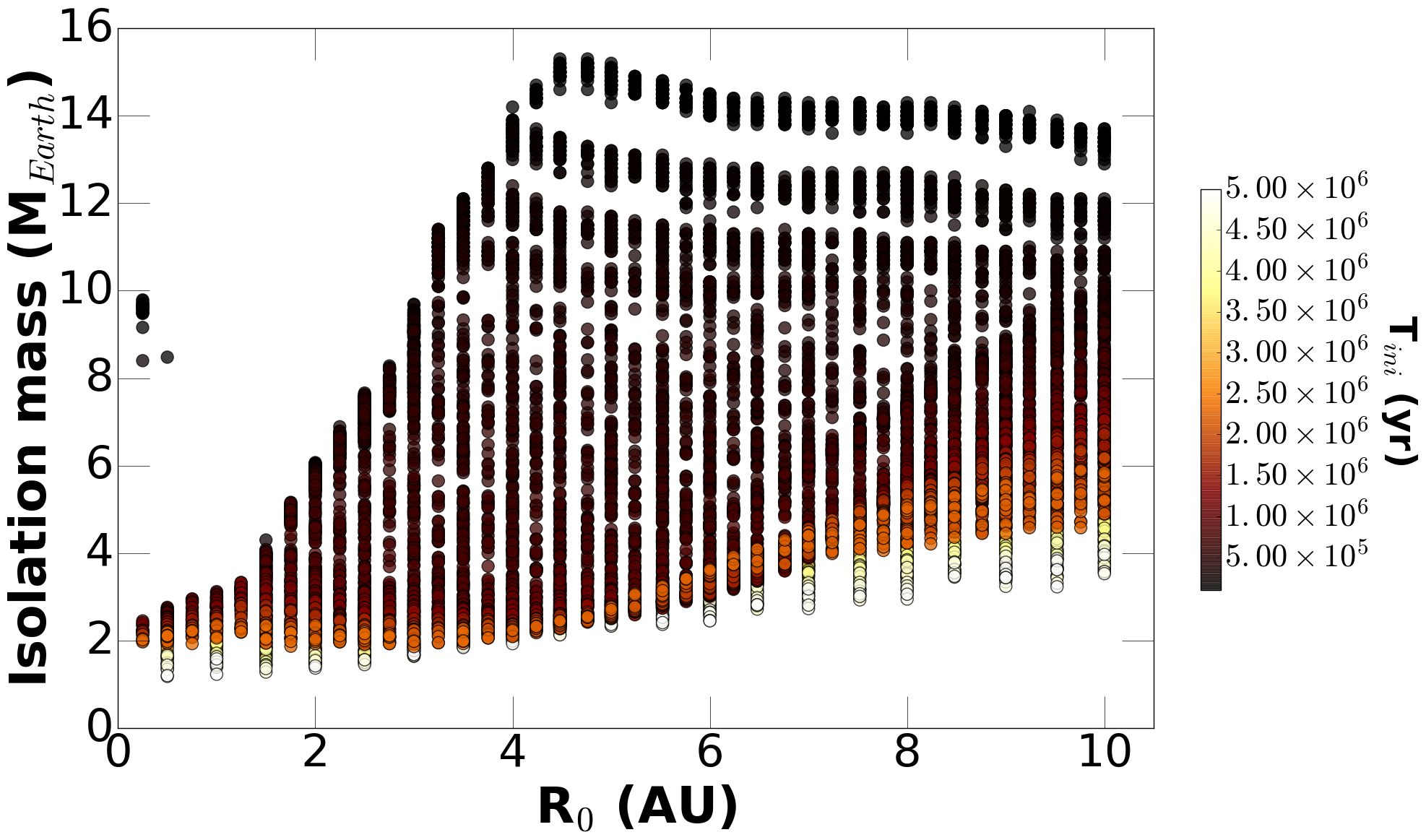}
   \caption{{The pebbles isolation mass reached by Neptune-mass planets as a function of their seed's injection time and location in the disk. The  pebbles isolation mass scales as {$(H/r)^3$}, and thus is higher in the outer regions of the disk, and for young disks. This plot however show the values actually reached by the simulated planet, and hence is affected by migration.}}
    \label{fig:mcoreall}
    \end{centering}
\end{figure*}


\begin{figure*}
\begin{centering}

	\includegraphics[scale=0.30]{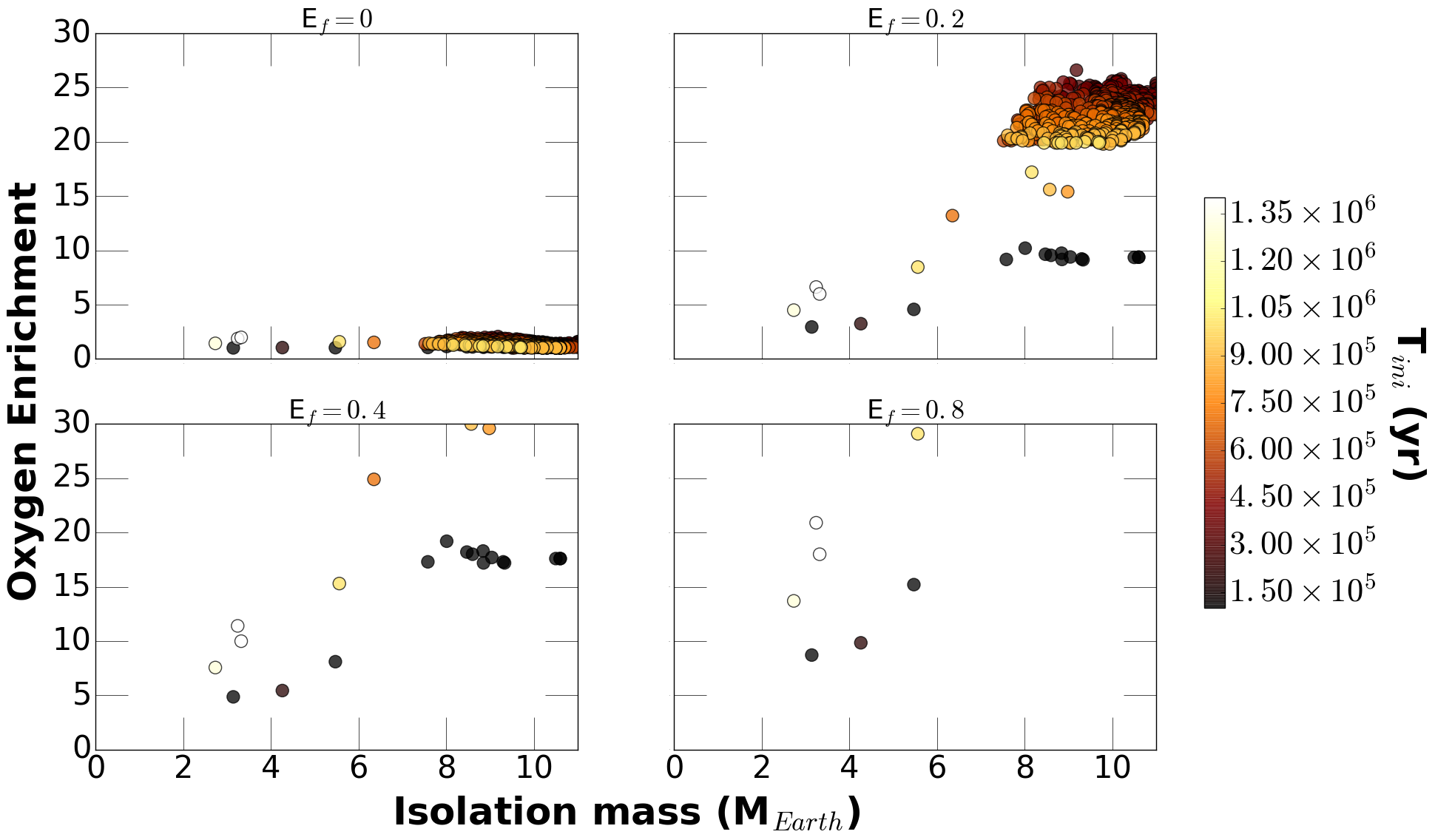}
   \caption{The oxygen enrichment as a function of the isolation mass and seed injection time of Neptune-mass planets found by the planetesimals accretion model, with 8 $\times$ MMSN metallicity and all of the HAT-P-26b constraints applied. This model fares better than the MMSN model, and favors a high isolation mass with low {enrichment factor}s.}
    \label{fig:plan2}
    \end{centering}
\end{figure*}

\begin{figure}
\begin{centering}
	\includegraphics[scale=0.55]{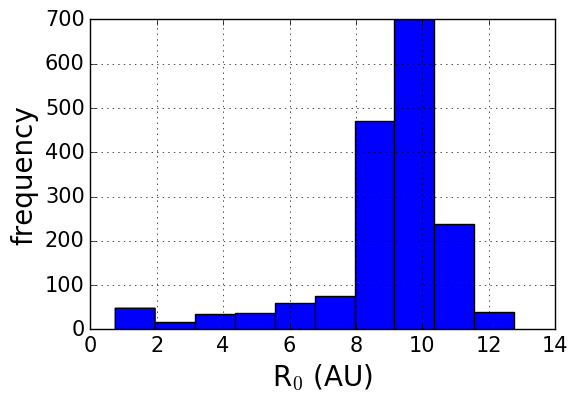}
   \caption{{The frequency of HAT-P-26b like planets as a function of their seed's initial injection location in the disk in the pebble accretion scenario. This frequency peaks between 8 and 11 AU, making this range the most probable region for the planet's initial formation phases.}}
    \label{fig:frequ}
    \end{centering}
\end{figure}



\section*{Acknowledgements}
Special thanks go to the Centre for Planetary Sciences group at the University of Toronto for useful discussions. We thank two anonymous referees for their constructive comments that significantly improved this manuscript.





\begin{thebibliography}{00}





\bibitem[\protect\citeauthoryear{Ali-Dib et al.}{2014a}]{ali-diba} Ali-Dib M., Mousis O., Petit J.-M., Lunine J.~I., 2014, ApJ, 793, 9 



\bibitem[\protect\citeauthoryear{Ali-Dib}{2017a}]{ali-dibc} Ali-Dib M., 2017, MNRAS, 464, 4282 
\bibitem[\protect\citeauthoryear{Ali-Dib}{2017b}]{ali-dibexo} Ali-Dib M., 2017, MNRAS, 467, 2845 

\bibitem[\protect\citeauthoryear{Ali-Dib, Johansen, \& Huang}{2017}]{ali-dibd} Ali-Dib M., Johansen A., Huang C.~X., 2017, MNRAS, 469, 5016 

\bibitem[\protect\citeauthoryear{Alibert}{2017}]{alibert} Alibert Y., 2017, arXiv, arXiv:1705.06008 


\bibitem[\protect\citeauthoryear{Baines et al.}{1995}]{baines} Baines K.~H., Mickelson M.~E., Larson L.~E., Ferguson D.~W., 1995, Icar, 114, 328 
\bibitem[\protect\citeauthoryear{Benz et al.}{2014}]{benz} Benz W., Ida S., Alibert Y., Lin D., Mordasini C., 2014, prpl.conf, 691 








\bibitem[\protect\citeauthoryear{Bitsch, Lambrechts, \& Johansen}{2015}]{bitsch1} Bitsch B., Lambrechts M., Johansen A., 2015, A\&A, 582, A112 


\bibitem[\protect\citeauthoryear{Cavali{\'e} et al.}{2017}]{cavalie} Cavali{\'e} T., Venot O., Selsis F., Hersant F., Hartogh P., Leconte J., 2017, Icar, 291, 1 

\bibitem[\protect\citeauthoryear{Dodson-Robinson \& Bodenheimer}{2010}]{robinson} Dodson-Robinson S.~E., Bodenheimer P., 2010, Icar, 207, 491 

\bibitem[\protect\citeauthoryear{Fortney, Marley, \& Barnes}{2007}]{fortney} Fortney J.~J., Marley M.~S., Barnes J.~W., 2007, ApJ, 659, 1661 

\bibitem[\protect\citeauthoryear{Fraine et al.}{2014}]{hat11} Fraine J., et al., 2014, Natur, 513, 526 

\bibitem[\protect\citeauthoryear{Guillot et al.}{1994}]{guillot1994} Guillot T., Gautier D., Chabrier G., Mosser B., 1994, Icar, 112, 337 

\bibitem[\protect\citeauthoryear{Guillot \& Gladman}{2000}]{gladman} Guillot T., Gladman B., 2000, ASPC, 219, 475 
\bibitem[\protect\citeauthoryear{Guillot et al.}{2004}]{guillot2004} Guillot T., Stevenson D.~J., Hubbard W.~B., Saumon D., 2004, jpsm.book, 1, 35 

\bibitem[\protect\citeauthoryear{Guillot \& Hueso}{2006}]{guillot2006} Guillot T., Hueso R., 2006, MNRAS, 367, L47 


\bibitem[\protect\citeauthoryear{Hartman et al.}{2011}]{hartman} Hartman J.~D., et al., 2011, ApJ, 728, 138 


\bibitem[\protect\citeauthoryear{Helled et al.}{2011}]{helledneptune} Helled R., Anderson J.~D., Podolak M., Schubert G., 2011, ApJ, 726, 15 

\bibitem[\protect\citeauthoryear{Helled \& Bodenheimer}{2014}]{helled} Helled R., Bodenheimer P., 2014, ApJ, 789, 69 


\bibitem[Karkoschka \& Tomasko(2011)]{karko} Karkoschka, E., \& Tomasko, M.~G.\ 2011, \icarus, 211, 780 

\bibitem[\protect\citeauthoryear{Lambrechts \& Johansen}{2012}]{lamb2012} Lambrechts M., Johansen A., 2012, A\&A, 544, A32 

\bibitem[\protect\citeauthoryear{Lambrechts \& Johansen}{2014}]{lamb1} Lambrechts M., Johansen A., 2014, A\&A, 572, A107 

\bibitem[\protect\citeauthoryear{Lambrechts, Johansen, \& Morbidelli}{2014}]{lamb2} Lambrechts M., Johansen A., Morbidelli A., 2014, A\&A, 572, A35 

\bibitem[\protect\citeauthoryear{Mordasini et al.}{2014}]{morda2014} Mordasini C., Klahr H., Alibert Y., Miller N., Henning T., 2014, A\&A, 566, A141 

\bibitem[\protect\citeauthoryear{Mousis et al.}{2014}]{mousis} Mousis O., Lunine J.~I., Fletcher L.~N., Mandt K.~E., Ali-Dib M., Gautier D., Atreya S., 2014, ApJ, 796, L28 

\bibitem[\protect\citeauthoryear{Movshovitz \& Podolak}{2008}]{mov} Movshovitz N., Podolak M., 2008, Icar, 194, 368 


\bibitem[\protect\citeauthoryear{Mullally et al.}{2015}]{mullaly} Mullally F., et al., 2015, ApJS, 217, 31 
\bibitem[\protect\citeauthoryear{Ormel}{2014}]{ormel2014} Ormel C.~W., 2014, ApJ, 789, L18 

\bibitem[\protect\citeauthoryear{Ormel, Shi, \& Kuiper}{2015}]{ormel} Ormel C.~W., Shi J.-M., Kuiper R., 2015, MNRAS, 447, 3512 

\bibitem[\protect\citeauthoryear{Podolak, Pollack, \& Reynolds}{1988}]{podolak} Podolak M., Pollack J.~B., Reynolds R.~T., 1988, Icar, 73, 163 
\bibitem[\protect\citeauthoryear{Podolak, Weizman, \& Marley}{1995}]{podolak1995} Podolak M., Weizman A., Marley M., 1995, P\&SS, 43, 1517 

\bibitem[\protect\citeauthoryear{Pollack et al.}{1996}]{pollack1996} Pollack J.~B., Hubickyj O., Bodenheimer P., Lissauer J.~J., Podolak M., Greenzweig Y., 1996, Icar, 124, 62 

\bibitem[\protect\citeauthoryear{Vazan et al.}{2016}]{vazan} Vazan A., Helled R., Podolak M., Kovetz A., 2016, ApJ, 829, 118 
\bibitem[\protect\citeauthoryear{Venturini, Alibert, \& Benz}{2016}]{vent} Venturini J., Alibert Y., Benz W., 2016, A\&A, 596, A90 

\bibitem[\protect\citeauthoryear{Wakeford et al.}{2017}]{hatp} Wakeford H.~R., et al., 2017, arXiv, arXiv:1705.04354 



\bibitem[\protect\citeauthoryear{Winn \& Fabrycky}{2015}]{winn} Winn J.~N., Fabrycky D.~C., 2015, ARA\&A, 53, 409 
\bibitem[\protect\citeauthoryear{Wilson \& Militzer}{2012}]{wilsonmilitzer} Wilson H.~F., Militzer B., 2012, ApJ, 745, 54 



\end{thebibliography}




\appendix


\bsp	
\label{lastpage}
\end{document}